\documentclass[vecphys]{svmult}
\usepackage{graphicx}
\usepackage{amssymb}
\usepackage[numbers]{natbib}

\makeatletter
 \usepackage{amsmath}

\usepackage{makeidx}         
\usepackage{multicol}        
\usepackage[bottom]{footmisc}
       
\makeindex             
\usepackage{aas_macros}
\usepackage{times}

\makeatother
\begin{document}

\title*{Matching the Local and Cosmic Star Formation Histories}

\titlerunning{Matching the Local and Cosmic SFHs}

\author{Igor Drozdovsky\inst{1,2}\and Andrew M. Hopkins\inst{3} \and
Antonio Aparicio\inst{1}\and Carme Gallart\inst{1}\and the
LCID team\inst{4}}

\authorrunning{Drozdovsky, I. et al.}

\institute{Instituto de Astrof\'{\i}sica de Canarias, C/V\'{\i}a L\'actea
s/n, 38200, La Laguna, Tenerife, Spain \texttt{dio@iac.es} \and Astronomical
Institute of St.Petersburg State University, Russia\and University
of Sydney, School of Physics, Bldg A28, NSW 2006, Australia \texttt{ahopkins@physics.usyd.edu.au}
\and \texttt{http://www.iac.es/project/LCID}}

\maketitle

\textbf{Given the many recent advances in our understanding of the
star formation history (SFH) of the Local Group and other nearby galaxies,
and in the evolution of star formation with redshift, we present a
new comparison of the comoving space density of the star formation
rate as a function of look-back time for the Local and Distant Universe.
We update the Local SFH derived from the analysis of resolved stellar
populations (``fossil records'') in individual nearby galaxies, based
on our own estimations as well as available in the literature. While
the preliminary comparison of SFHs is found to be broadly consistent,
some discrepancies still remain, including an excess of the Local
SFR density in the most recent epoch.} 

The goal of this project is to establish whether the star formation
history (SFH) of galaxies in the local Universe is consistent with
the cosmic (global) SFH inferred from surveys of distant, `high-redshift',
galaxies.  A common approach based on redshift surveys aims to measure
indicators of recent star formation in galaxies at different distances,
which due to the finite speed of light, give us a view of different
cosmic epochs. One of the drawbacks of this method is that it is not
directly possible to connect different galaxies, measured at different
redshifts, into a coherent evolutionary sequence, to provide a consistent
picture of \emph{observed} galaxy evolution. Moreover, the detail
in which these galaxies can be studied is limited since they are mostly
unresolved, and their faintness limits the wavelength resolution of
any spectroscopic measurements. An alternative and complementary approach
focuses on galaxies near enough to be resolved into their component
stars. For these systems we can use well established stellar evolution
theory, together with photometry and spectroscopy of individual stars
of various ages, to interpret the {}``fossil record\char`\"{} of
their star formation, and to trace the evolution of each from its
formation to the present time. Despite a limited volume and sample
size, the high quality SFHs one can obtain with this method is valuable
in a general cosmological context. In particular, it provides important
clues to the earliest stages of star formation that are progressively
harder to probe with galaxy redshift surveys targeting the most distant
systems, as well as to probe issues that are not accessible through
large-scale galaxy surveys (e.g., uncertainties in conversion factors
from luminosity to SFR, dust attenuation, shape of the stellar and
galaxy mass functions, etc). Initial comparisons suggest these different
approaches do not yield the same results \citep{Tolstoy:98,Hopkins:01},
but the errors are large, paradoxically, due to the lack of complete
normalized SFHs of nearby galaxies. 

New space- and ground-based multi-wavelength wide-field observational
data, better understanding of stellar structure and evolution, and
the recent development, coupled with greater computing power, of automated
techniques that measure the star formation histories by statistical
comparisons of models and observations \citep[e.g.,][]{Tolstoy:96,Aparicio:96,Dolphin:97,Hernandez:99,Harris:01}
have radically modified our view of star formation in the Milky Way
(MW) and Andromeda (M31), the two gravitationally dominant galaxies
of the Local Group \citep[e.g.,][]{Williams:03,Brown:06}. Our knowledge
of the star formation history in other nearby galaxies has also been
significantly improved, thanks to the possibility of analysing their
faint and ancient resolved stellar populations\citep[e.g., ][]{Dolphin+05,Gallart+07}.
These low mass stars have extremely long lifetimes, comparable to
the age of the Universe, and retain in their atmospheres the gas,
with the elemental abundances intact, from the time of their birth.
Thus they provide critical information on the early (initial) star
formation rate and heavy element abundance. The total number of galaxies
where these old ($>10$\,Gyr) stars are measurable has doubled since the
2001 publication of \citet{Hopkins:01}.

In the past few years the measurement of the evolution of star formation
rate (SFR) in various types of galaxies at a broad range of redshifts
has also tremendously progressed. An example of an extensive compilation
drawn from the literature of SFR density measurements at redshifts
$0<z<6$ has been done by \citet{Hopkins:06}, following an investigation
that carefully addressed issues associated with corrections for obscuration
by dust within galaxies, internally consistent calibrations for the
various star-formation rate indicators, and constraints on the initial
distribution of stellar masses in a burst of star formation, among
others. Taken together, the most recent and robust data indicate a
compellingly consistent picture of the cosmic SFH constrained to within
factors of about $\sim3$. There is now growing evidence that the
evolution is essentially flat beyond $z\sim1$, however it is still
unclear whether at $z>3$ the evolution flattens, declines or continues
to increase.

\begin{figure}
\includegraphics[width=11.5cm,keepaspectratio]{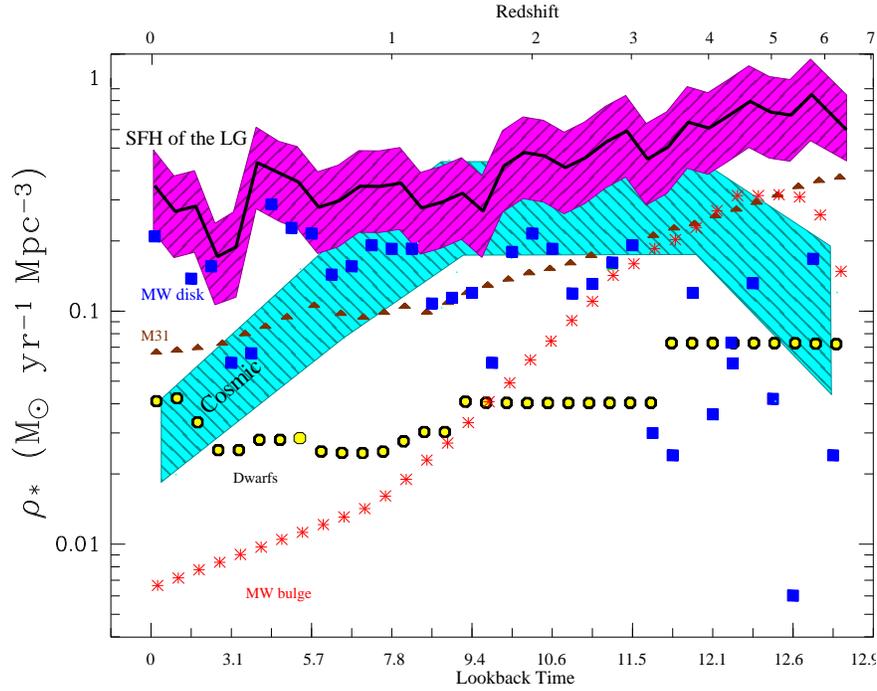} 

\caption{Comparison of the Local Group SFH (solid line and magenta-shaded
area) with that from a compilation of redshift surveys (cyan-shaded
area) for a $(H_{0},\Omega_{M},\Omega_{\Lambda})=(70,0.3,0.7)$ cosmology.
The squares, stars, triangles and circles show the contributions from
the MW disk, MW bulge, M31 and dwarf galaxies respectively. \label{fig:MatchingSFHs}}
\end{figure}

On Fig.~\ref{fig:MatchingSFHs}, we present an updated comparison
of the comoving space density of the star formation rate as a function
of look-back time for the Local Group (LG) and distant Universe. To
measure the comoving space density for the LG, we integrated the best
currently known SFHs for the galaxies within a volume defined by a
1.5~Mpc-radius sphere, each normalized by its current total star
formation rate, similar to the \citet{Hopkins:01} approach. For the
MW and M31 we combined various estimations of SFH for each of galactic
stellar components, such as disk, bulge, and halo.

The cosmic SFH (cyan shaded area) is taken from the compilation of
\citet{Hopkins:06}, and scaled to ensure consistency with the Salpeter
(1955) initial mass function. The shaded areas indicate the uncertainty
in the star formation rate densities for the LG (dominated by  uncertainties
in the total star formation histories of the MW and M31), and for
the cosmic star formation rate. 

The results of this comparison are: 

\begin{itemize}
\item An excess of the local star formation density in the recent $\sim5$~Gyr
is mainly due to the fluctuations of star formation of the MW disk.
\item Between $\sim8$ and $\sim12$~Gyr, the SFH of the LG is broadly
consistent with the Cosmic one.
\item The early/initial evolution of the LG was dominated by the spheroidal
component of the MW and M31.
\item The overall trend of $\rho_{*}$ from the LG supports a fairly flat
evolution of the SFR, suggesting factor of $\sim10$ extinction correction
to high-redshift, UV-based SFR measures.
\end{itemize}

\begin{figure}
\includegraphics[width=12cm,keepaspectratio]{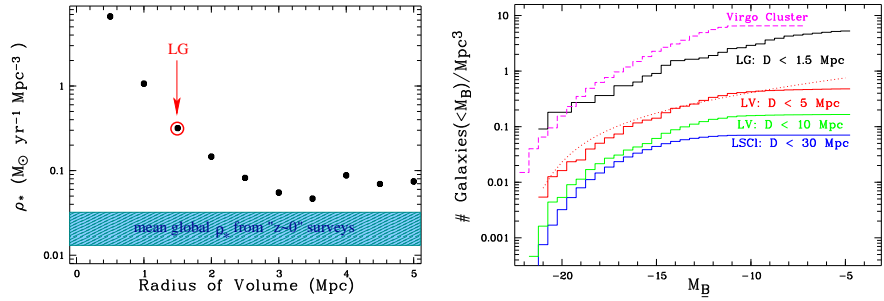}

\caption{\emph{Left:} Current (H$\alpha$) star formation density as a function
of volume size. \emph{Right panel:} Comulative luminosity function
(B band) for the different samples of galaxies within 30~Mpc.\label{fig:propvsvolume} }
\end{figure}

The important question is whether star formation in the Local Group
was representative of the cosmic mean. Following an extension of the
copernican principle, one can assume that the Local Group has no special
location in the Universe and hence its SFHs should be explained rather
naturally, without appeal to special conditions. Certainly there are
some pecularities of the LG environment from the larger volume, such
as a dominance by two large spirals, an excess of the local current
SFR density, and a deficiency of dwarf galaxies in comparison with
denser environments (Fig.~\ref{fig:propvsvolume}; Drozdovsky et
al. 2008, in prep.). At the same time, about 85\% of agalaxies are
situated outside the dense environments of rich clusters, with roughly
half of them belonging to groups like the Local Group, and the remaining
half scattered in the `field' \citep{ikar:05,BHawthorn+06}. An extension
of the ``fossil record'' studies to a larger volume and improving
our knowledge of the LG SFH will provide a more robust comparison
of the Local and Cosmic evolution. 


\end{document}